\documentstyle[twoside,fleqn,espcrc2,epsf]{article}

\def\psca{$\nu e^+ \to \nu e^+$ }
\def\prod{$\nu \to \nu e^- e^+$ }

\newcommand{\AmS}{{\protect\the\textfont2
  A\kern-.1667em\lower.5ex\hbox{M}\kern-.125emS}}

\title{Plasma influence on the neutrino - electron processes 
       in a strong magnetic field}

\author{A.V. Kuznetsov and N.V. Mikheev\address{
Division of Theoretical Physics, Department of Physics, \\
Yaroslavl State University, Sovietskaya 14, \\
150000 Yaroslavl, Russian Federation}
        \thanks{Supported by INTAS grant N~96-0659 and in part 
by Russian Foundation for Basic Research Grant N~98-02-16694.
A.K. thanks the organizers for the hospitality.} 
}
       
\begin{document}

\setcounter{page}{0}

\thispagestyle{empty}

{\large
\begin{flushright}
{\normalsize Yaroslavl State University\\
             Preprint YARU-HE-99/05\\
             hep-ph/9907493} \\[10mm]
\end{flushright}

\begin{center}
{\bf Plasma influence on the neutrino - electron processes \\
       in a strong magnetic field}\\[10mm]
{\bf A.V. Kuznetsov and N.V. Mikheev} \\[5mm]
Division of Theoretical Physics, Department of Physics, \\
Yaroslavl State University, Sovietskaya 14, \\
150000 Yaroslavl, Russian Federation
\end{center}

\vspace{20mm}

An influence of the magnetized electron - positron plasma on the absorption 
and loss of the energy and momentum in a process 
of neutrino propagation is investigated. A total contribution of all crossed 
processes, $\nu \to \nu e^- e^+$, $\nu e^- \to \nu e^-$, $\nu e^+ \to \nu 
e^+$, $\nu e^- e^+ \to \nu$, is found for the first time, which appears not to 
depend on the chemical potential of electron-positron gas. 
Relatively simple expressions for the probability and mean losses of the 
neutrino energy and momentum are obtained, which are suitable for 
a quantitative analysis. 

\vspace{20mm}

\begin{center}
{\it Based on the talks presented at \\
the Xth International Baksan School ``Particles and Cosmology'', \\
Baksan Valley, Kabardino Balkaria, Russia, April 19-25, 1999 \\
and\\
the International Workshop ``Particles in Astrophysics and Cosmology: \\
From Theory to Observation'', Valencia, Spain, May 3-8, 1999}
\end{center}
}

\newpage
\maketitle

\begin{abstract}
An influence of the magnetized electron - positron plasma on the absorption 
and loss of the energy and momentum in a process 
of neutrino propagation is investigated. A total contribution of all crossed 
processes, $\nu \to \nu e^- e^+$, $\nu e^- \to \nu e^-$, $\nu e^+ \to \nu 
e^+$, $\nu e^- e^+ \to \nu$, is found for the first time, which appears not to 
depend on the chemical potential of electron-positron gas. 
Relatively simple expressions for the probability and mean losses of the 
neutrino energy and momentum are obtained, which are suitable for 
a quantitative analysis. 
\end{abstract}

\section{Strong magnetic fields in astrophysics}

An understanding of the important role of neutrino interactions in 
astrophysical processes stimulates a constantly growing interest in the 
neutrino physics in a dense medium~\cite{Raff}. 
Matter is usually considered as the 'medium'. It should be stressed, that
a strong magnetic field can play the role of the active medium.
However, the magnetic field influences significantly the quantum processes 
only in the case when it is strong enough. 
There exists a natural scale for the field strength which is the so-called 
critical Schwinger value $B_e = m_e^2/e \simeq 4.41 \cdot 10^{13}$ G
(we use natural units in which $c = \hbar = 1$). 

The fields of such strength are unattainable in a laboratory. 
However, the astrophysical objects and processes 
inside them give us unique possibilities for investigations of the 
particle physics, and of the neutrino physics especially 
under extreme conditions of a strong magnetic field. A concept of the 
astrophysically strong magnetic field has been changed in the recent years 
(see fig. 1). 
Whereas the magnetic fields with the strength $10^9 \div 10^{11}$ G 
were considered as 'very strong' near thirty years ago,  
the fields have been observed at the surface of pulsars have appeared to 
be stronger, of the order of $10^{12} \div 10^{13}$ G. 
These fields are now treated as 
the so-called `old' magnetic fields. The fields at a moment of the cataclysm
like a supernova explosion, when a neutron star was born, could be much 
greater.
In the present view, the magnetic field strength in the astrophysical 
processes like a supernova explosion or a coalescence of 
neutron stars could be as high as $\sim 10^{15} \div 10^{17}$ G. 
The possible existence of such fields both of toroidal~\cite{Bis,tor}, 
and of poloidal~\cite{pol} types is the subject of wide speculation. 

\begin{figure*}[htb]

\epsfbox[97 480 525 690]{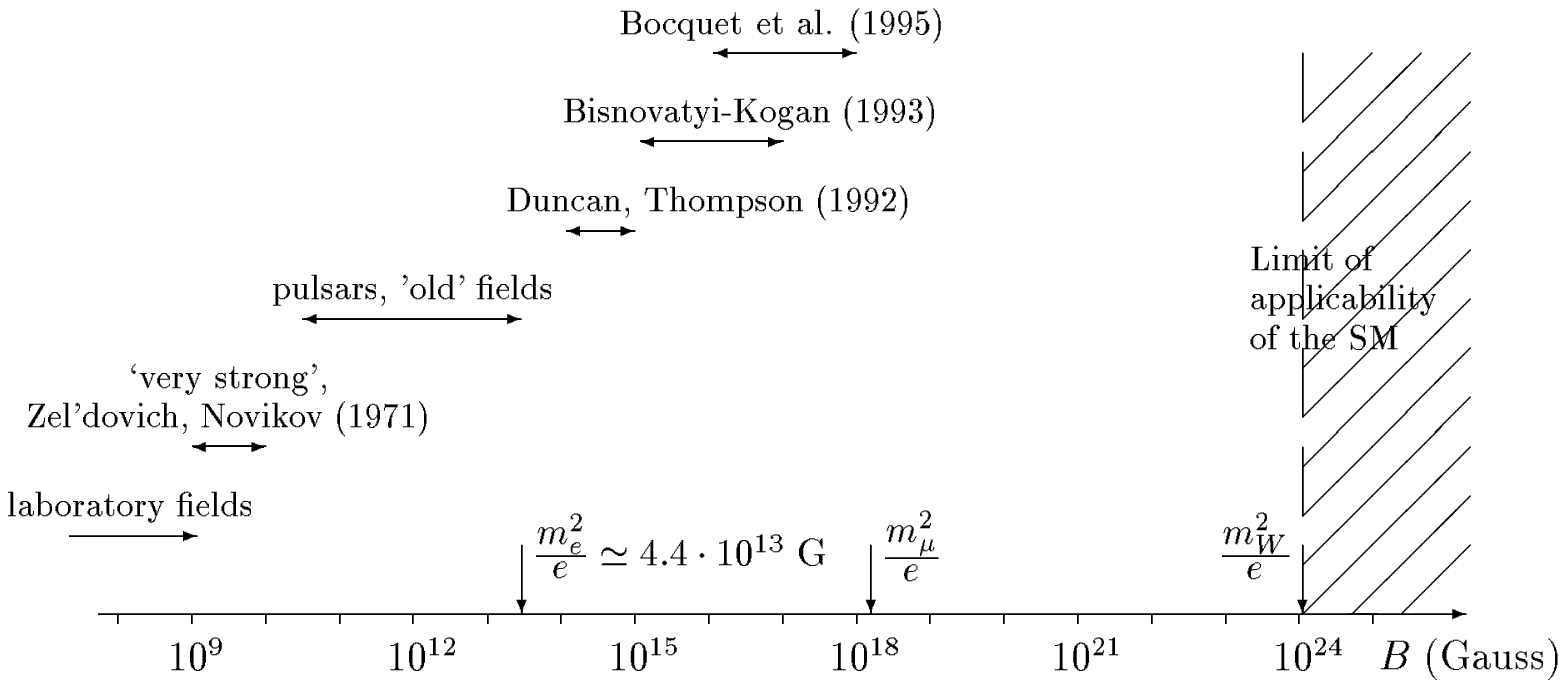}

\caption{Evolution of the notion 'strong magnetic field' in astrophysics.} 
\label{fig:mag-scale}

\end{figure*}

The mechanism of a generation of such strong magnetic field of the toroidal 
type suggested by 
Bisnovatyi-Kogan many years ago~\cite{Bis} looks as the most convincing one. 
Really, a question should be asked of how natural is it to expect 
an existence of strong magnetic fields in astrophysics? 

i) What can be considered as the exotic case: a star with magnetic field or 
a star without it? It is believed that a star without magnetic field is rather 
the exotic object. It is also believed that it is more natural 
for a star to have the primordial magnetic field before the stage of collapse. 
The primordial field at the level of 100 G readily leads 
in the compression to the fields $\sim 10^{12}$ G. 

ii) What can be considered as more natural case: a rotating star or a 
non-rotating one? 
It is believed that a non-rotating star is rather the exotic object. 

iii) What kind of collapse looks more exotic: a compression as a solid body or 
a compression with the gradient of the angular velocity? 
It is believed that a compression without the gradient is rather the exotic 
case. 

The above three points are necessary for a realization of the 
Bisnovatyi-Kogan mechanism. The field lines, by virtue of the angular velocity
gradient, become twisted and more dense. As the analysis shows, the fields 
could be winded from $\sim 10^{12}$ G to $\sim 10^{16}$ G during a few 
seconds. 

Thus the appearance of strong magnetic fields of the toroidal type 
$\sim 10^{16}$ G in the stellar collapse looks more natural than the 
absense of such fields.

It should be emphasized that the field of the order of $10^{16} $G 
is really a rather dense medium with the mass density 
\begin{equation}
\rho \, = \, \frac{B^2}{8 \pi} \, \simeq \, 0.4 \cdot 10^{10} 
{\mbox{g} \over \mbox{cm}^3} \cdot 
\left({B \over {10^{16} \mbox{G}}}\right)^2 ,
\label{eq:rho}
\end{equation}

\noindent 
which is comparable with the plasma mass density $10^{10} - 10^{12} 
\mbox{g/cm}^3$ to be typical for the envelope of an exploding supernova. 

\section{Neutrino - electron processes in a strong magnetic field}

Neutrino does not interact with a magnetic field directly, except for models 
where it has an unusually large magnetic moment. The field influence on a 
neutrino is provided by quantum processes of its weak interaction with 
charged fermions. Electron plays here the main role as the particle with 
the maximal specific charge $e / m_e$. 

If the momentum transferred is relatively small, $|q^2| \ll m^2_W$, the 
weak interaction of neutrinos with electrons could be described in the local 
limit by the effective Lagrangian of the following form (for additional 
conditions of applicability of the local limit in a strong field see below)
\begin{equation}
{\cal L} \, = \, \frac{G_F}{\sqrt 2} 
\big [ \bar e \gamma_{\alpha} (C_V - C_A \gamma_5) e \big ] \,
\big [ \bar \nu \gamma^{\alpha} (1 - \gamma_5) \nu \big ] \,,
\label{eq:L}
\end{equation}

\noindent 
where $C_V = \pm 1/2 + 2 sin^2 \theta_W, \, C_A = \pm 1/2$.
Here the upper signs correspond to the electron neutrino 
($\nu = \nu_e$) when both $Z$ and $W$ boson exchange contributes to 
the interaction. The lower signs correspond to $\mu$ and $\tau$ neutrinos 
($\nu = \nu_{\mu}, \nu_{\tau}$), when the $Z$ boson exchange 
is only presented in the Lagrangian~(\ref{eq:L}). 

Two parameters are essential in the analysis of definite neutrino processes 
in a magnetic field, which are the field strength and the 
neutrino energy. Thus, two asymptotic 
limits exist where calculations become much easier.

i) The `weak' field limit.

We call the field as a `weak' one in the case $e B \ll E_\nu^2$. 
This condition can be rewritten in the Lorentz invariant form. 
The presence of the particle four-momentum $p^\alpha = (E, {\bf p})$ 
allows, in parallel with the field invariant
\begin{equation}
e^2 F_{\mu \nu} F^{\nu \mu} \equiv e^2 (F F) = - 2 e^2 B^2, 
\label{eq:inv1}
\end{equation}

\noindent 
to construct the dynamical invariant
\begin{equation}
e^2 p_\mu F^{\mu \nu} F_{\nu \rho} p^\rho \equiv e^2 (p F F p) 
= e^2 B^2 E_\nu^2 \sin^2 \theta, 
\label{eq:inv2}
\end{equation}

\noindent 
where $\theta$ is an angle between the particle momentum
$\bf p$ and the magnetic field induction $\bf B$. 
The invariant~(\ref{eq:inv2}) is most commonly used in the dimensionless form
\begin{equation}
\chi^2 = e^2 (p F F p)/m^6 .
\label{eq:hi2}
\end{equation}

\noindent Thus the condition of the field `weakness' appears
\begin{equation}
\left[e^2 (F F)\right]^{3/2} \ll e^2 (p F F p) .
\label{eq:cond1}
\end{equation}

\noindent In this case the electrons occupy the higher Landau levels. 
It is well-known that one can use the crossed field approximation, when 
the invariant $(F F) = 0$, in the investigations of neutrino processes 
in the limit~(\ref{eq:cond1}). 

Returning to the local Lagrangian~(\ref{eq:L}) we could formulate the 
conditions of its applicability in a strong magnetic field as follows: 
the $W$ boson mass should exceed 
essentially the field parameters~(\ref{eq:inv1}) and~(\ref{eq:inv2}) 
\begin{equation}
m_W^2 \gg \left[e^2 (F F)\right]^{1/2}, \,\left[e^2 (p F F p)\right]^{1/3},
\label{eq:condloc}
\end{equation}

\noindent 
which yields correspondingly
$e B \ll m_W^2$, 

\noindent 
$E \ll m_W^3 / e B$. 

ii) The `strong' field limit.

In this limit the field strength $B$ appears to be the largest physical 
parameter, $e B \gg E^2$, or in the invariant form
\begin{equation}
\left[e^2 (F F)\right]^{3/2} \gg e^2 (p F F p) .
\label{eq:cond2}
\end{equation}

\noindent In this case the electrons occupy only the lowest Landau level. 

Most likely the first investigation of the neutrino - electron process 
induced by a magnetic field was performed for the so-called synchrotron 
emission of neutrino pairs by electrons, $e \to e \nu \bar\nu$~\cite{Baier} 
which is kinematically forbidden in vacuum.
The crossed process to the neutrino pair emission is the production of the 
electron - positron pair by a neutrino, \prod, which is also kinematically 
opened only in the presence of a magnetic field. The probability of 
this process in a `weak' field was calculated in the 
papers~\cite{Choban,Bor}. In our papers \cite{KM97}
the process of the production of the $e^+ e^-$ pair by a neutrino was 
investigated in both cases of the `weak'~(\ref{eq:cond1}) and 
`strong'~(\ref{eq:cond2}) field limits.  
The process probability was calculated and the four-vector connected with 
the mean values of the neutrino energy and momentum loss was found 
\begin{equation}
Q^{\alpha} \, = - \, 
E \left (\frac{d E}{d t}, \frac{d {\bf p}}{d t} \right )
 \, = \, E \int (p - p')^\alpha d W, 
\label{eq:Qal}
\end{equation}

\noindent where $p$ and $p'$ are the momenta of the initial and final 
neutrinos correspondingly, 
$d W$ is the total differential probability of the process per unit time. 

In the two above-mentioned asymptotic limits we obtained the four-vector 
$Q^{\alpha}$ in the following form

i) $e B \gg E^2$
\begin{eqnarray}
&&\hspace{-7mm}Q^{\alpha} = \frac{G_F^2 e B (p \varphi \varphi p)^2}{48 \pi^3} 
\label{eq:Q1} \\
&&\hspace{-7mm}\times \{(C_V^2 + C_A^2)[p^{\alpha} 
- 2 (\varphi \varphi p)^{\alpha}] 
+ 2 C_V C_A (\tilde \varphi p)^{\alpha} \} ,
\nonumber 
\end{eqnarray}

\noindent 
where $\varphi_{\alpha \beta}$ is the dimensionless external field tensor,
$\varphi_{\alpha \beta} = F_{\alpha \beta}/B$,
$\tilde \varphi_{\alpha \beta} = {1 \over 2} \varepsilon_{\alpha \beta \rho 
\sigma} \varphi^{\rho \sigma}$ is the dual tensor; 

ii) $e B \ll E^2$
\begin{eqnarray}
&&\hspace{-7mm}Q^{\alpha} \, = \, {7 \over 16} \, 
\frac{G_F^2 (C_V^2 + C_A^2)}{27 \pi^3} \, 
m^6 \chi^2 \, 
\nonumber \\
&&\hspace{-7mm}\times [\,p^{\alpha} (ln \chi - 1.888) 
- \sqrt{3} \, \frac{\beta^2}{\chi} \,
(\varphi \varphi p)^{\alpha} 
\label{eq:Q2} \\
&&\hspace{-7mm}- 7.465 \, \frac{C_V C_A}{C_V^2 + C_A^2} \, \frac{\beta}{\chi^{2/3}} \,
(\tilde \varphi p)^{\alpha}],
\nonumber 
\end{eqnarray}

\noindent 
where $\beta = B/B_e$. 

This four vector defines the energetic and force action on the medium by a 
neutrino. Really, after integration over the initial neutrino 
distribution it defines the energy delivered to the unit volume 
per unit time, and the volume density of the neutrino force acting 
on the medium. 

\section{Neutrino - electron processes in strongly magnetized plasma} 

We have discussed up to now the processes in a pure magnetic field of a high 
intensity. 
However, in the most of astrophysical objects both components of the 
active medium, a magnetic field and plasma, are likely to be presented. 
For example, the astrophysical cataclysms are known like 
a supernova explosion or a coalescence of neutron stars where rather dense 
electron - positron plasma exists on the periphery, and strong magnetic 
fields of the toroidal type can be generated, $B \sim 10^{16}$ G, 
due to the Bisnovatyi-Kogan mechanism. 
The mass density of such field, see eq.~(\ref{eq:rho}), exceeds 
essentially the density of the electron component of plasma on the 
periphery, where the total density is of order of $10^{12} \mbox{g/cm}^3$.  
Thus the magnetic field is the dominating factor, $e B \gg T^2, \mu^2$, 
where $\mu$ is the chemical potential of electrons,
$T$ is the temperature of plasma. The last condition can be formulated more 
carefully if one compares the energy densities of the magnetic field 
and the magnetized electron - positron plasma 
\begin{equation}
\frac{B^2}{8 \pi} \, \gg  \, \frac{\pi^2 (n_{e^-} - n_{e^+})^2}{e B} 
+ \frac{e B T^2}{12}, 
\label{eq:energy}
\end{equation}

On the other hand, rather high neutrino energies are typical for 
astrophysical processes, $E_\nu, T \gg m_e$. Thus we shall consider 
the physical situation when the field strength appears to be the 
largest parameter, while the electron mass is the smallest one
\begin{equation}
e B \gg E_\nu^2, \mu^2, T^2 \gg m_e^2.
\label{eq:cond3}
\end{equation}

All the neutrino - electron processes described by the 
Lagrangian~(\ref{eq:L}) can be separated into two parts:

i) the processes with the neutrino - antineutrino pair in the initial or 
in the final state, $e^- e^+ \leftrightarrow \nu \bar\nu$, 
$e \leftrightarrow e \nu \bar\nu$;

ii) the processes where the neutrino presents both in the initial and 
in the final state, $\nu e^\mp \leftrightarrow \nu e^\mp$, 
$\nu \leftrightarrow \nu e^- e^+$, and the similar processes with the 
antineutrino. 

A simple analysis shows that in the case of a strong field, 
see~(\ref{eq:cond2}) and~(\ref{eq:energy}), when the electrons occupy 
only the lowest Landau level, the processes with the neutrino pair are 
strongly suppressed. Really, the total spin of the neutrino - antineutrino 
pair in the center-of-mass system is equal to 1, while 
the total spin of the electron - positron pair on the lowest Landau level 
is equal to zero. Thus the process amplitude is strictly zero for the case of 
massless particles, and it is suppressed in the considered relativistic limit. 
Furthermore, the process of the synchrotron emission 
of the neutrino pair, $e \to e \nu \bar\nu$, just as the reverse process, 
are forbidden by the energy and momentum conservation when the initial and 
the final electrons are on the lowest Landau level. 

Among the other processes which accompany the neutrino propagation in a 
magnetic field and plasma, the `canonical' scattering processes 
$\nu e^- \to \nu e^-$, \psca 
are possible without the presence of the magnetic field. The 
analysis of the $\nu e$ scattering in the magnetized plasma was 
performed in the paper~\cite{Bez}. 
The results of numerical calculations of the paper~\cite{Bez} 
show that the magnetic field influence on the total scattering cross-section 
is insignificant 
in the range of parameters considered which corresponds in fact 
to the 'weak' field limit~(\ref{eq:cond1}).  

Taking account of the plasma influence on the indicated `exotic' process, 
\prod, is reduced to a simple modification by insertion of the statistical 
factor for the electron and the positron in the final state. 
The second `exotic' process of the absorption of the pair, 
$\nu e^- e^+ \to \nu$, becomes 
possible only in the presence of a magnetic field and plasma 
simultaneously. 

The probability of this process in a unit time has a physical 
meaning only being integrated over the initial electron and positron states 
\begin{eqnarray}
&&\hspace{-7mm}W (\nu e^- e^+ \to \nu)
\label{eq:Wdef} \\
&&\hspace{-7mm} = \frac{1}{\cal T} \int |{\cal S}|^2 
\; d \Gamma_{e^-} \; f_{e^-} 
\; d \Gamma_{e^+} \; f_{e^+} 
\; d \Gamma'_\nu \; (1 - f'_\nu), 
\nonumber 
\end{eqnarray}

\noindent 
where $\cal T$ is the total interaction time, 
$d \Gamma$ is the phase-space element of a particle, 
$f$ is its distribution function, 
$f'_\nu = [exp((E'-\mu_\nu)/T_\nu) + 1]^{-1}$, 
$\mu_\nu, T_\nu$ are the chemical potential and the temperature of 
the neutrino gas, 
$f_{e^{\mp}} = [exp((E_{\mp} \mp \mu)/T) + 1]^{-1}$, 
$\mu, T$ are the chemical potential and the temperature of the 
elec\-tron-po\-si\-tron gas. 
The $\cal S$ matrix element of the transition coinsides with the one for the 
process $\nu \to \nu e^- e^+$~\cite{KM97}, to the crossing transformation. 
We do not present here the details of integration over the phase space of 
particles, which will be published in an extended paper. The result of our 
calculation of the probability~(\ref{eq:Wdef}) can be presented in the form 
of a single integral
\begin{eqnarray}
&&\hspace{-7mm} W (\nu e^- e^+ \to \nu) = \frac{G_F^2 e B T^2 E}{4 \pi^3} 
\label{eq:Wpick} \\
&&\hspace{-7mm}\times [(g_V + g_A)^2 (1 - u)^2 + (g_V - g_A)^2 (1 + u)^2]
\nonumber\\
&&\hspace{-7mm}\times \int\limits_0^\infty 
\frac{d \xi}{(e^\xi - 1) (1 + e^{-x + \eta_\nu - \xi/\tau})} 
\, \ln{\frac{\cosh{\xi} + \cosh{\eta}}{1 + \cosh{\eta}}},
\nonumber
\end{eqnarray}

\noindent 
where $x = E/T_\nu$, $\tau = T_\nu/T$, 
$\eta_\nu = \mu_\nu/T_\nu$, $\eta = \mu/T$, $u = \cos \theta$, 
$\theta$ is an angle between the initial neutrino momentum
$\bf p$ and the magnetic field induction $\bf B$. 

The probabilities of the remaining crossed processes 
$\nu e^- \to \nu e^-$, $\nu e^+ \to \nu e^+$, 
$\nu \to \nu e^- e^+$ are defined similarly to~(\ref{eq:Wdef}) 
with the substitution $f_{e^\mp} \to (1 - f_{e^\pm})$ in every transposition 
of the electron (positron) from the initial state to the final one. 
For the total probability of the neutrino interaction with magnetized 
elec\-tron-po\-si\-tron plasma
\begin{eqnarray}
&&\hspace{-7mm} W (\nu \longrightarrow \nu) 
\nonumber\\
&&\hspace{-7mm} 
= W (\nu \to \nu e^- e^+) + W(\nu e^- e^+ \to \nu) +
\nonumber\\
&&\hspace{-7mm} + W (\nu e^- \to \nu e^-) + W (\nu e^+ \to \nu e^+), 
\label{eq:Wtotd}
\end{eqnarray}

\noindent 
we obtain
\begin{eqnarray}
&&\hspace{-7mm} W (\nu \longrightarrow \nu)  = 
\frac{G_F^2 e B T^2 E}{4 \pi^3} 
\bigg \lbrace 
(g_V + g_A)^2 (1 - u)^2 
\nonumber\\
&&\hspace{-7mm} \times 
\int\limits_0^{x \tau \frac{1 + u}{2}} 
\frac{\xi d \xi}{(1 - e^{-\xi}) (1 + e^{-x + \eta_\nu + \xi/\tau})} \,
\nonumber\\
&&\hspace{-7mm} + (g_V - g_A)^2 (1 + u)^2 
\label{eq:Wtot}\\
&&\hspace{-7mm} \times 
\int\limits_0^{x \tau \frac{1 - u}{2}} 
\frac{\xi d \xi}{(1 - e^{-\xi}) (1 + e^{-x + \eta_\nu + \xi/\tau})} \,
\nonumber\\
&&\hspace{-7mm} + 
[ (g_V + g_A)^2 (1 - u)^2 + (g_V - g_A)^2 (1 + u)^2]
\nonumber\\
&&\hspace{-7mm} \times 
\int\limits_0^\infty 
\frac{\xi d \xi}{(e^\xi - 1) (1 + e^{-x + \eta_\nu - \xi/\tau})}
 \bigg \rbrace .
\nonumber
\end{eqnarray}

\noindent 
It is interesting to note that the dependence on the electron chemical 
potential exactly cancelled in the total probability~(\ref{eq:Wtot}), whereas 
each of the partial probabilities~(\ref{eq:Wtotd}) does depend on $\mu$. 
We do not know a physical underlying reason of this cancellation up to now. 
Probably, some property of a completeness of the considered set of processes 
with respect to the electrons manifests itself here. 

The probability~(\ref{eq:Wtot}) defines the partial contribution of the 
considered processes into the neutrino opacity of the medium. 
The estimation of the neutrino mean free path with respect to the 
neu\-tri\-no-elec\-tron processes yields
\begin{equation}
\lambda_e = {1 \over W} \simeq 170 \;\mbox{km} \cdot
\left ({{10^3 B_e} \over B} \right ) \;
\left ({{5\;\mbox{MeV} \over T}} \right )^3.
\label{eq:lambda}
\end{equation}

\noindent 
It should be compared with the mean free path caused by the interaction 
with nuclei, which is evaluated to be of order of 1 km at the density 
value $\rho \sim 10^{12} \;\mbox{g/cm}^3$. At first glance the influence 
of the neu\-tri\-no-elec\-tron reactions on the process of neutrino 
propagation is negligibly 
small. However, a mean free path does not exhaust the neutrino physics in 
a medium. The mean values of the neutrino energy and momentum loss are more 
essential in astrophysical applications, and especially the asymmetry of 
the momentum loss, caused by the influence of an external magnetic field. 
Many attempts were made to calculate such asymmetry due to 
neu\-tri\-no-nu\-cle\-on processes, motivated by the problem of the observed 
high space velocities of pulsars (see~\cite{DongLai} and references therein). 
As the analysis shows, the contribution of the neu\-tri\-no-elec\-tron 
processes into the asymmetry could be comparable with the contributions of the 
neu\-tri\-no-nu\-cle\-on processes. 

The components of the four-vector of losses $Q^{\alpha}$~(\ref{eq:Qal}) 
should be calculated from the formula
\begin{equation}
Q^{\alpha} \, = \, E \int q^\alpha d W, 
\label{eq:Qdef}
\end{equation}

\noindent 
where $q$ is the difference of the momenta of the initial and final neutrinos, 
$q = p - p'$, $d W$ is the total differential probability of the processes 
presented in Eq.~(\ref{eq:Wtotd}). The result of our calculation of the 
zeroth and third components (the magnetic field is directed along the third 
axis) of the four-vector $Q^{\alpha}$ is
\begin{eqnarray}
&&\hspace{-7mm} Q_{0,3}  =  
\frac{G_F^2 e B T^3 E^2}{4 \pi^3} \;
\nonumber\\
&&\hspace{-7mm} \times
\bigg \lbrace (g_V + g_A)^2 (1 - u)^2 
\nonumber\\
&&\hspace{-7mm} \times
\int\limits_0^{x \tau \frac{1 + u}{2}} 
\frac{\xi^2 d \xi}{(1 - e^{-\xi}) 
(1 + e^{-x + \eta_\nu + \xi/\tau})} \,
\nonumber\\
&&\hspace{-7mm} \pm
(g_V - g_A)^2 (1 + u)^2 
\label{eq:Q03p}\\
&&\hspace{-7mm} \times
\int\limits_0^{x \tau \frac{1 - u}{2}} 
\frac{\xi^2 d \xi}{(1 - e^{-\xi}) 
(1 + e^{-x + \eta_\nu + \xi/\tau})} \,
\nonumber\\
&&\hspace{-7mm} -
[ (g_V + g_A)^2 (1 - u)^2 \pm (g_V - g_A)^2 (1 + u)^2]
\nonumber\\
&&\hspace{-7mm} \times
\int\limits_0^\infty 
\frac{\xi^2 d \xi}{(e^\xi - 1) 
(1 + e^{-x + \eta_\nu - \xi/\tau})}
 \bigg \rbrace .
\nonumber
\end{eqnarray}

\noindent 
Our results for the probability and the four-vector of losses obtained in 
the case of a pure magnetic field~\cite{KM97}, are reproduced from 
Eqs.~(\ref{eq:Wtot}) and~(\ref{eq:Q03p}) in the limit of a rarefied 
plasma ($T,\, T_\nu,\, \mu_\nu \to 0$). 

\section{Possible astrophysical manifestations} 

To illustrate the results obtained we estimate the volume density of the 
neutrino energy loss per unit time, $\dot{\cal E}$, and the volume density of the 
neutrino force acting on plasma along the magnetic field, 
${\cal F}$
\begin{equation}
(\dot{\cal E}, {\cal F}) 
= \int d n_\nu \, {1 \over E} Q_{0,3},
\label{eq:EFdef}
\end{equation}

\noindent 
where $d n_\nu$ is the neutrino density 
\begin{equation}
d n_\nu = {d^3 p \over (2 \pi)^3} \,\,{1 \over 
{e^{{E - \mu_\nu} \over T_\nu} + 1}} .
\label{eq:dn}
\end{equation}

\noindent 
The case $T_\nu = T$ corresponds to the neutrino equilibrium distribution. 
For the values defined in Eq.~(\ref{eq:EFdef}) we obtain
\begin{eqnarray}
&&\hspace{-7mm} (\dot{\cal E}, {\cal F}) = 
(g_V^2 + g_A^2, 2 g_V g_A) \frac{G_F^2 e B T_\nu^7}{3 \pi^5} 
\label{eq:EFres} \\
&&\hspace{-7mm} \times
\int\limits_0^\infty \frac{x^3 d x}{e^{x - \eta_\nu} + 1}
\int\limits_0^\infty 
\frac{y^2 d y}{1 + e^{-x - y + \eta_\nu}}
\frac{e^{(\tau - 1) y} - 1}{e^{\tau y} - 1} .
\nonumber
\end{eqnarray}

\noindent 
The formula~(\ref{eq:EFres}) demonstrates that the neutrino action on plasma 
tends to zero in a case when the neutrino distribution is 
the exact equilibrium one, $\tau = T_\nu/T \to 1$. 
We stress that the origin of the force density ${\cal F}$ in 
Eq.~(\ref{eq:EFres}) is the interference of the vector and axial-vector 
couplings in the effective Lagrangian~(\ref{eq:L}), and it appears as the 
macroscopic manifestation of the parity violation in weak interaction.  
At first glance, the main contribution into ${\cal F}$ seems to arise from 
the electron neutrinos, because $g_V (\nu_e) \gg g_V (\nu_{\mu,\tau})$. 
However, as will be shown later, the dominant contribution comes from the 
muon and tau neutrino and antineutrino. 

For the sake of numerical estimations it is convenient to present 
Eq.~(\ref{eq:EFres}) in the following form 
\begin{equation}
(\dot{\cal E}, {\cal F}) \simeq {\cal A}\;
(g_V^2 + g_A^2, 2 g_V g_A) \,\varphi (\eta_\nu) \,
\psi (\tau) ,
\label{eq:EFapp}
\end{equation}

\noindent 
where
\begin{eqnarray}
&&\hspace{-7mm} {\cal A} = \frac{12 G_F^2 e B T^7}{\pi^5} 
\label{eq:A}\\
&&\hspace{-7mm} 
= \left({B \over {10^{16} \mbox{G}}}\right)
\left ({T \over {4\;\mbox{MeV}}} \right )^7 
\cases{0.55 \cdot 10^{20}\;{\mbox{dyne} \over \mbox{cm}^3},\cr \cr
       1.6 \cdot 10^{30}\;{\mbox{erg} \over \mbox{cm}^3 \cdot\mbox{s}},\cr}
\nonumber\\
\nonumber\\
&&\hspace{-7mm} 
\varphi (\eta_\nu) = {\eta_\nu^4 \over 24} +
{\pi^2 \eta_\nu^2 \over 12} +
{7 \pi^4 \over 360} + \mbox{Li}_4 (- e^{-\eta_\nu}),
\label{eq:phi}\\
&&\hspace{-7mm} 
\varphi (0) = {7 \pi^4 \over 720} \simeq 0.947,
\nonumber\\
&&\hspace{-7mm} 
\psi (\tau) = {\tau^7 \over 6} \, \int\limits_0^\infty 
\frac{y^2 d y}{e^{\tau y} - 1} \left [e^{(\tau - 1) y} - 1 \right ], 
\label{eq:psi} \\
&&\hspace{-7mm} 
\psi (\tau)\big\vert_{\tau \to 1} \simeq \tau - 1 , 
\nonumber
\end{eqnarray}

\noindent 
$\mbox{Li}_4 (z)$ is the polylogarythm function. The formulas obtained are 
also valid for the processes with antineutrino due to the $CP$ invariance 
of the weak interaction. 

The neutrino distribution in a supernova envelope is known to deviate 
from the equilibrium. Neutrinos outgoing from the central part of a star, 
having a high temperature, enter the periphery region where the 
strong magnetic field
is generated and where the temperature of the elec\-tron-po\-si\-tron gas 
is lower. 
The spectral temperatures are known to be different for the neutrinos of 
various types~\cite{Raff}, $T_{\nu_e} < T_{\bar\nu_e} < 
T_{\nu_{\mu,\tau}} \simeq T_{\bar\nu_{\mu,\tau}}$. 
The neutrino action on plasma leads to the establishment of the thermal 
equilibrium, $\dot{\cal E}_{tot} = 0$. 
In an analysis of the equilibrium the contributions into $\dot{\cal E}_{tot}$ 
of all processes of the neutrino interaction with matter should be taken into 
account. The $\beta$ processes $\nu_e + n \leftrightarrow e^- + p$ are known, 
see e.g.~\cite{Raff}, to dominate in the energy balance. The rate of the 
plasma heating due to these processes can be presented in the form 
$\dot{\cal E} (\beta) \simeq {\cal B} (T_{\nu_e} - T)/T$. As indicated above, 
the $\beta$ process probability is much greater than the one for the 
neutrino - electron processes, therefore it is natural that ${\cal B} \gg 
{\cal A}$, where ${\cal A}$ is defined in Eq.~(\ref{eq:A}). Consequently, 
the plasma temperature has to be established very close from above to the 
electron neutrino spectral temperature ($T \simeq T_{\nu_e}, \; 
T > T_{\nu_e}$). 
It is important to note that the force density ${\cal F} \ne 0$ 
when $\dot{\cal E}_{tot} = 0$. 
Taking for the estimation $T_{\nu_e} = 4 \,\mbox{MeV}$, 
$T_{\bar\nu_e} = 5 \,\mbox{MeV}$, 
$T_{\nu_{\mu,\tau}} = T_{\bar\nu_{\mu,\tau}} = 8 \,\mbox{MeV}$,
and considering the chemical potentials to be small, we found for the force 
density 
\begin{equation}
{\cal F} \simeq 
3.6 \cdot 10^{20}\;{\mbox{dyne} \over \mbox{cm}^3} \cdot
\left({B \over {10^{16} \mbox{G}}}\right).
\label{eq:F}
\end{equation}

\noindent 
Strictly speaking, the angular asymmetry should be also included into 
the neutrino distribution~(\ref{eq:dn}). However, as the analysis shows, 
it could change our result~(\ref{eq:F}) at no more than 10 \%. 

The force~(\ref{eq:F}) should be compared with the recent calculation of 
the similar force caused by the $\beta$ processes~\cite{GvOg}. 
Under the same physical conditions our result appears to be of the same 
sign and is larger than the result of Ref.~\cite{GvOg} by a factor of 2 
or more. 
Thus the role of the neutrino - electron processes in a strong magnetic 
field could appear more essential than the $\beta$ processes. 
There is good reason to believe that the results obtained could be useful 
in a detailed theoretical description of the process of 
supernova explosion. 

{\bf Acknowledgements}  

We are grateful to G.~Raffelt and V.~Semikoz for helpful discussions.


\begin{thebibliography}{99}
%
\bibitem{Raff} 
   G.G.~Raffelt, Stars as Laboratories for Fundamental
   Physics, University of Chicago Press, Chicago, 1996.
%
\bibitem{Bis}
   G.S.~Bisnovatyi-Kogan, Physical Questions of a Theory of the Star 
   Evolution, Nauka Publ., Moscow, 1989 (in Russian); 
   G.S.~Bisnovatyi-Kogan and S.G.~Moiseenko, Astro. Zh. 69 (1992) 563 
   [Sov. Astron. 36 (1992) 285];
   G.S.~Bisnovatyi-Kogan, Astron. Astrophys. Transactions 3 (1993) 287.
%
\bibitem{tor} 
   M.~Ruderman, in ``Neutron Stars: Theory and Observation'', ed. by 
   J.~Ventura and D.~Pines, Kluwer Academic. Pub., Dordrecht, 1991; 
   G.J.~Mathews et al., preprint astro-ph/9710229 on xxx.lanl.gov.
%
\bibitem{pol}
   R.C.~Duncan and C.~Thompson, Astrophys.J., 392 (1992) L9; 
   C.~Thompson and R.C.~Duncan, Mon.Not.R.Astron.Soc., 275 (1995) 255; 
   M.~Bocquet at al., Astron. and Astrophys., 301 (1995) 757.
%
\bibitem{Baier} 
   V.N.~Baier, V.M.~Katkov, Dokl. Akad. nauk SSSR 171 (1966) 313.
%
\bibitem{Choban} 
   E.A.~Choban, A.N.~Ivanov, ZhETF 56 (1969) 194.
%
\bibitem{Bor}
   A.V.~Borisov, V.Ch.~Zhukovskii, B.A.~Lysov, Izv. Vuz. Fiz. 8 (1983) 30 
   [Sov. Phys. J. 26 (1983) 701];
   A.V.~Borisov, A.I.~Ternov, V.Ch.~Zhukovsky, Phys. Lett. B 318 (1993) 489. 
%
\bibitem{KM97}
   A.V.~Kuznetsov, N.V.~Mikheev, Phys. Lett. B 394 (1997) 123; 
   Yad. Fiz. 60 (1997) 2038 [Phys. At. Nucl. 60 (1997) 1865].
%
\bibitem{Bez}
   V.G.~Bezchastnov, P.~Haensel, Phys. Rev. D 54 (1996) 3706.
%
\bibitem{DongLai} 
   P.~Arras, D.~Lai, E-print astro-ph/9811371. 
%
\bibitem{GvOg}
   A.A.~Gvozdev, I.S.~Ognev, Pis'ma v ZhETF. 69 (1999) 337 
   [JETP Lett. 69 (1999) 365].
\end{thebibliography}
\end{document}